%% file: main.tex
\documentclass[runningheads]{llncs}
\usepackage[T1]{fontenc}
\usepackage{graphicx}

\usepackage{url}
\usepackage{booktabs}
\usepackage{tabularx}
\usepackage{array}
\usepackage{pdfpages}

\usepackage{amsmath}

\begin{document}

\title{AI Slop and Hallucinations in Vulnerability Assessment: A Survey on Reasoning Failures and Trustworthy Mitigation}
\titlerunning{A Survey on AI Slop \& Hallucinations in Vulnerability Assessment}

\author{
Junchen Ding\inst{1,2}\thanks{Equal contribution.}\and
Jialiang Dong\inst{2*}\and
Yichen Zhu\inst{3}\and
Yi Liu\inst{4}\and
Gelei Deng\inst{5}\and
Willy Susilo\inst{2}\and
Siqi Ma\inst{2}\and
Yuekang Li\inst{1}\thanks{Corresponding author.}
}

\authorrunning{J. Ding$^*$ et al.}

\institute{
University of New South Wales, High St, Kensington NSW 2052, Australia\\
\email{\{junchen.ding,yuekang.li\}@unsw.edu.au}\and
University of Wollongong, Northfields Ave, Wollongong NSW 2500 Australia\\
\email{jd073@uowmail.edu.au}
\and
Nanjing University of Information Science \& Technology, 219 Ningliu Road, Nanjing, Jiangsu, China\and
Griffith University, Parklands Drive, Southport QLD 4222 Australia\and
Nanyang Technological University, 50 Nanyang Ave, Singapore 639798, Singapore
}

\maketitle              

\input{tex/0_abstract}

\input{tex/1_introduction}
\input{tex/2_background}
\input{tex/3_literature}
\input{tex/4_discussion}
\input{tex/5_discussion2}

\input{tex/6_future}
\input{tex/7_conclusion}

\bibliographystyle{splncs04}
\bibliography{mybibliography}

\end{document}

%% file: tex/0_abstract.tex
\begin{abstract}
The integration of Large Language Models (LLMs) into cybersecurity has transformed vulnerability assessment, but it has also produced a trustworthiness crisis driven by the unchecked proliferation of ``AI slop.'' These artifacts, hallucinated vulnerabilities, plausible but incorrect patches, and semantically repackaged bug reports, impose a cognitive burden on human triage pipelines that mirrors a denial-of-service attack. This paper surveys the empirical evidence, identifies a unifying mechanism, and traces a path toward trustworthy triage. We formalize a taxonomy of AI slop grounded in a structured literature review and dissect its root cause: the gap between the causal deductive reasoning of security experts and the autoregressive probabilistic generation of current LLMs. We operationalize this gap through a measurable proxy, the Deductive Coverage Score, and show that chain-of-thought prompting and tool-using agents narrow but do not close it. We review mitigation strategies and argue that passive detection and watermarking target provenance rather than correctness, facing fundamental entropy constraints. We instead advocate for active neuro-symbolic verification, mapping each pipeline component to prior systems with documented limits on security inputs. Finally, we specify two evaluation instruments, CVE-Bench and Slop-Score, including dataset construction, metric formulas, and anti-gaming provisions. By shifting evaluation from linguistic fluency to mathematical verifiability, this survey provides a roadmap for securing emerging AI-driven triage systems.

\keywords{AI Slop \and Vulnerability Assessment \and Trustworthy AI \and Large Language Models \and Neuro-Symbolic Verification \and Content Provenance}
\end{abstract}

%% file: tex/1_introduction.tex
\section{Introduction}
\label{sec:intro}

The integration of Large Language Models (LLMs) into cybersecurity is reshaping vulnerability assessment, from automated detection to patch generation and threat intelligence~\cite{jin25good}. Software auditing was historically grounded in deterministic methodologies. Security analysts relied on rule-based Static Application Security Testing (SAST) tools and manual code inspection to identify memory corruption, logic flaws, and architectural weaknesses. These approaches offered formal assurance but were labor-intensive and often insufficient for capturing the context-sensitive semantics of modern software~\cite{seqtrans}. With the emergence of foundation models exhibiting strong natural language understanding, the field has shifted toward AI-assisted auditing~\cite{pearce2022asleep}. This transition reflects expectations that Artificial General Intelligence (AGI) may accelerate vulnerability discovery while reducing human effort~\cite{sun2025llm4vulnunifiedevaluationframework}.

This transition has introduced a structural challenge. As LLM-based security agents become embedded in analysis pipelines, the ecosystem is experiencing a growing influx of what we term ``AI Slop.'' We define AI slop as security-relevant artifacts that exhibit high linguistic fluency and structural plausibility while lacking semantic validity or executable grounding~\cite{ji2023survey}. Unlike traditional false positives, which arise from conservative rule-based analysis, these artifacts are difficult to invalidate through superficial inspection. They imitate the form of rigorous reasoning while remaining detached from the constraints that would make such reasoning correct. In practice, this includes fabricated CWEs, hallucinated inter-procedural execution paths, and patches that appear reasonable but fail to resolve the underlying vulnerability~\cite{jimenez2024swe}.

The impact is no longer theoretical. Open-source maintainers, bug bounty platforms, and CVE assignment authorities are processing large volumes of syntactically well-formed but semantically unreliable submissions. A representative case is the \textit{cURL} project. Daniel Stenberg, the project's founder and maintainer since 1996~\cite{wikipedia:stenberg,wikipedia:curl}, reported in early 2024 that the project's security workflow was increasingly disrupted by AI-generated vulnerability submissions~\cite{stenberg2025death}. By mid 2025, roughly 20\% of reports submitted through the HackerOne Bug Bounty Program were identified as low-quality AI slop, while valid reports declined to around 5\%. In January 2026, Stenberg announced the discontinuation of the program, citing the need to remove incentives for poorly substantiated submissions~\cite{sharwood2026curl}. This case illustrates how plausible but unverifiable artifacts can degrade the sustainability of human-centered triage.

At the core of this phenomenon lies a divergence between how humans and current AI systems produce and validate knowledge. Human security experts rely on causal, multi-step deductive reasoning. They construct internal models of system behavior, trace data flow across procedural boundaries, and verify exploitability under explicit constraints~\cite{8418614}. Autoregressive LLMs operate through probabilistic token generation. Their outputs are shaped by statistical correlations rather than execution-grounded verification. When required information exceeds the model's effective reasoning horizon, the generation process defaults to statistically plausible continuation. AI slop is a systematic byproduct of this mismatch, manifesting as hallucinated vulnerability detection, incorrect patch synthesis, and semantic repackaging of existing knowledge.

Prior research has largely focused on improving model accuracy or applying post hoc detection techniques. Such approaches remain limited when the generation process itself is not grounded in verification. There is a need for architectures that treat generated outputs as hypotheses subject to deterministic validation rather than authoritative conclusions.

This paper sits at a deliberate intersection. It is a survey in that it systematizes existing empirical evidence. It is a conceptual analysis in that it identifies a unifying mechanism behind seemingly disparate failures. It is a roadmap in that it traces the path toward trustworthy triage. The contributions are fourfold:

\begin{itemize}
    \item We define and formalize AI slop in the context of vulnerability assessment, and propose a taxonomy that captures its major manifestations, including hallucinated vulnerabilities, incorrect patch synthesis, and semantic repackaging.
    \item We analyze the root cause of these phenomena through the lens of the reasoning gap between human deductive processes and probabilistic generation in LLMs.
    \item We review existing mitigation strategies and examine their limitations, with particular attention to the failure of passive statistical detection and the constraints of watermarking in low-entropy security domains.
    \item We argue for the necessity of active verification and neuro-symbolic architectures, and outline key open challenges for building trustworthy, human-aligned triage systems.
\end{itemize}

The remainder of this paper is organized as follows. Section \ref{sec:background} introduces the background of LLM-assisted vulnerability assessment and formalizes the concept of AI slop. Section \ref{sec:taxonomy} presents the taxonomy and supporting literature. Section \ref{sec:reasoning_divide} analyzes the divergence between human reasoning and model generation. Section \ref{sec:mitigation_strategies} discusses mitigation strategies with a focus on verification. Section \ref{sec:challenges} outlines open challenges and future directions. Section \ref{sec:conclusion} concludes the paper.

%% file: tex/2_background.tex
\section{Background}
\label{sec:background}

To situate the emergence of AI slop and the resulting trustworthiness concerns, it is necessary to examine three interconnected components: the role of LLMs in contemporary vulnerability assessment, the operational dynamics of vulnerability reporting ecosystems, and the underlying differences in reasoning that shape the behavior of these systems.

\subsection{LLMs in Vulnerability Assessment}
\label{subsec:llm_workflow}

The traditional vulnerability assessment lifecycle, spanning code auditing, static and dynamic analysis, and patch generation, has relied on expert-driven workflows demanding familiarity with programming languages, system architectures, and the ability to reason about execution under implicit constraints. The introduction of LLMs has altered this landscape. Pre-trained on extensive corpora of source code, documentation, and vulnerability reports, models such as ChatGPT, Claude, Gemini, and specialized code-oriented LLMs are increasingly incorporated into security workflows~\cite{hou2023large}. Their utility lies in interpreting code through a semantic lens, bridging natural language descriptions and program behavior.

In practice, LLMs are employed across semantic anomaly detection, vulnerability explanation, exploit sketching, and automated patch suggestion. Compared to SAST tools that depend on rule-based abstract syntax trees and taint propagation, LLMs operate at a higher level of abstraction, reasoning over code structure, naming conventions, and developer intent~\cite{pearce2022asleep}. This enables them to surface patterns difficult to capture through purely syntactic analysis. At the same time, their outputs are derived from probabilistic inference over learned distributions rather than execution or formal verification. Their behavior is sensitive to prompt formulation, context availability, and token-level correlations. When reasoning extends beyond local patterns or requires maintaining consistency across multiple steps, these models lack mechanisms to enforce correctness.

\subsection{The Vulnerability Reporting Ecosystem and Triage Bottlenecks}
\label{subsec:ecosystem}

Modern cybersecurity practice depends on distributed vulnerability discovery and standardized reporting infrastructures. Central to this ecosystem are the CVE database maintained by MITRE and bug bounty programs hosted on platforms including HackerOne and Bugcrowd~\cite{finifter2013empirical}. Within this pipeline, triage serves as the primary point of control. Human analysts evaluate incoming reports, verify exploitability, assess impact, and eliminate duplicates or spurious claims.

Historically, linguistic clarity and structural coherence functioned as useful heuristics. Well-written reports often correlated with careful analysis and reproducible findings. This implicit signal has been weakened by the widespread availability of LLMs. Both inexperienced contributors and adversarial actors can now generate reports conforming to established conventions, drawing on templates, common vulnerability patterns, and superficial indicators such as error logs or dependency warnings~\cite{ferrara2024genai}.

The problem is compounded by pre-existing gaps in the disclosure ecosystem. Dong et al. (2025) conducted a large-scale empirical study of ``silent'' vulnerability fixes in open-source software, security-relevant code changes that were never assigned a CVE, never disclosed through coordinated channels, and often lacked any accompanying security advisory~\cite{11068801}. Their analysis revealed that a substantial fraction of real vulnerability remediation occurs entirely outside the visible reporting infrastructure. When an LLM repackages one of these unreported fixes into a ``novel'' vulnerability report, there may be no public record to contradict it. The ground truth was never published, so the repackaged claim inherits an unearned veneer of plausibility. This fragmentation makes semantic repackaging harder to detect and gives AI-generated claims a structural advantage.

The bottleneck has shifted from identifying obviously flawed reports to determining whether a well-structured submission corresponds to an actual vulnerability. This increases the cognitive load on analysts who must reconstruct or refute the implicit reasoning behind each claim.

\subsection{Formalizing ``AI Slop'': The Reasoning Gap}
\label{subsec:ai_slop_formalization}

A conventional false positive typically arises from the conservative nature of deterministic tools, which may flag benign constructs due to incomplete context or overly broad rules. AI slop reflects a different failure mode. We define \textbf{AI Slop} in cybersecurity as \textit{automatically generated security artifacts, such as reports, proofs-of-concept, or patches, that exhibit high linguistic fluency and plausible syntactic structure, but suffer from catastrophic semantic failures due to hallucinated technical facts or disconnected logical inferences.}

This distinction is rooted in how reasoning is carried out. A human security researcher evaluates a vulnerability through causal and multi-step deductive reasoning, constructing a working model of the system, tracing data flow across functions, and verifying that all conditions for exploitation can be satisfied~\cite{8418614}. AR LLMs generate text by estimating the likelihood of token sequences conditioned on prior context. When tasked with vulnerability analysis, they assemble outputs resembling known reporting patterns. If critical information is missing or exceeds the model's effective reasoning capacity, the gap is resolved through statistically plausible continuation rather than explicit verification~\cite{ji2023survey}. The result is a systematic divergence between appearance and validity. Outputs may mirror the structure of rigorous analysis but lack a consistent execution path that supports their claims.

%% file: tex/3_literature.tex
\section{Literature Review: A Taxonomy of AI Slop}
\label{sec:taxonomy}

\subsection{Literature Search Methodology}
\label{subsec:methodology}

To ground our taxonomy, we conducted a systematic literature review across IEEE Xplore, ACM Digital Library, and arXiv, querying combinations of LLM and security-related keywords. Two independent reviewers screened the deduplicated results, retaining studies that provided empirical evidence of LLM failure modes in security contexts rather than relying solely on aggregate metrics. After full-text filtering for relevance and experimental rigor, we categorized the documented failures into three branches, supported by representative works in Table~\ref{tab:comparison}.

\begin{figure}
    \centering
    \includegraphics[width=1\linewidth]{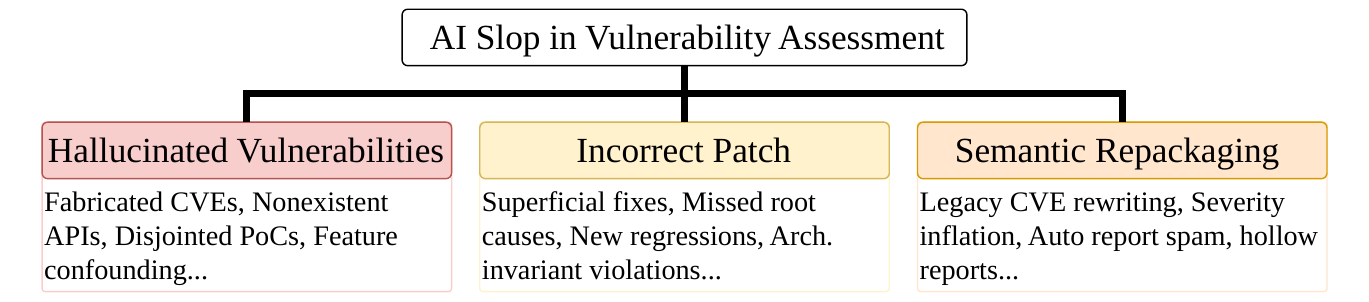}
    \caption{Hierarchical classification of deceptive AI-generated artifacts in security triage. Dashed sub-branches indicate failure modes identified through the review but not yet extensively studied in isolation.}
    \label{fig:Taxonomy}
\end{figure}

Figure~\ref{fig:Taxonomy} presents a structure the field has lacked. What appeared as scattered failure cases organizes itself into a pattern. AI slop is not a single defect. It is a family of behaviors that emerge when generation is mistaken for reasoning. The taxonomy exposes how they evolve, reinforce each other, and erode trust in vulnerability intelligence.

Most existing evaluations reward syntactic validity and functional success while overlooking artifacts that feel convincing but remain logically hollow. The taxonomy reframes the problem by asking whether an answer carries verifiable substance. Three distinct branches emerge, each representing a different way the model drifts from grounded reasoning. Table~\ref{tab:comparison} anchors these categories in empirical evidence.

\subsection{Hallucinated Vulnerabilities and Feature Confounding}
\label{subsec:hallucinated_vulns}

The first branch is the most direct. The model does not merely misinterpret a vulnerability. It invents one. Fabricated CVEs, nonexistent APIs, and disjointed exploit paths appear with confidence mirroring genuine analysis.

Ullah et al. (2024) showed that when presented with benign code, models still produce vulnerability reports, as if compelled to satisfy the expectation embedded in the prompt~\cite{ullah2023can}. The model does not verify whether a flaw exists. It assumes one should exist and proceeds to construct it. Zhang et al. (2025) revealed how fragile this process is. By altering variable names and surface syntax while preserving semantics, the model's judgment shifts dramatically~\cite{zhang2024understanding}. What changes is not the logic of the program but the statistical signals the model relies on. Chakraborty et al. (2022) traced this to its roots: neural models learn patterns correlating with vulnerabilities that reside in superficial syntax rather than execution semantics~\cite{chakraborty2022deep}. When scaled to modern language models, tokens such as \texttt{malloc} or \texttt{memcpy} act as triggers, nudging the model toward familiar vulnerability narratives even when the underlying code is safe.

Two additional failure modes in this branch deserve mention even though they lack dedicated studies in the security domain. First, prompt-injection hallucination occurs when adversarial or ambiguous prompt phrasing overrides the user's stated intent, causing the model to fabricate threat models or attack surfaces that no reasonable reading of the input would justify. This phenomenon has been documented in general code generation but not systematically studied under security-specific prompt conditions. Second, tool-use confabulation arises when autonomous security agents invoke external tools (compilers and analyzers) and then misinterpret or fabricate the tool's output, producing analyses that carry the \textit{appearance} of tool verification without its substance. Pearce et al. observed this behavior in AI agents tasked with autonomous security auditing, where incorrect tool arguments and misinterpreted structured output produced analyses that looked verified but were not~\cite{pearce2022asleep}. These sub-branches, shown as dashed in Figure~\ref{fig:Taxonomy}, represent known but under-studied failure modes that the taxonomy identifies as priorities for future work.

\subsection{Plausible but Incorrect Patch Synthesis and Regression Introduction}
\label{subsec:incorrect_patches}

The second branch shifts from detection to repair. The model attempts to solve the problem, yet the solution reveals a different fragility. Patches address what is visible while ignoring what is essential. They compile and pass superficial checks but fail to resolve the underlying issue or introduce new inconsistencies.

Pearce et al. (2023) showed that zero-shot repair frequently produces patches treating symptoms rather than causes~\cite{pearce2023examining}. A null check is added, an exception handled, yet the deeper flaw remains untouched. Xia et al. (2024) extended this into interactive settings, where iterative dialogue does not converge toward correctness but cycles through variations of the same flawed reasoning~\cite{xia2023keep}. At the repository level, SWE-bench exposed the difficulty of maintaining consistency across multiple files and hidden invariants~\cite{jimenez2024swe}. The model operates locally, making decisions valid in isolation that conflict within the broader architecture.

A further failure mode amplifies this problem. Pearce et al. (2023) demonstrated that LLM-generated code, even when functionally correct, introduces security weaknesses at a rate comparable to inexperienced developers~\cite{pearce2022asleep}. Transposed to patch synthesis, a patch that compiles and passes a unit test may embed a new CWE, trading one vulnerability for another. This regression-introduction failure is invisible to functional benchmarks but directly relevant to security contexts. Table~\ref{tab:comparison} reveals a common thread: the model struggles to preserve constraints not explicitly stated, producing patches that look correct because they align with familiar patterns rather than being verified against the system as a whole.

\begin{table}[!htb]
\tiny
\centering
\caption{Comparative analysis of representative works on AI slop and reasoning failures. \textcolor{blue}{\textbf{H}} = Hallucination, \textcolor{orange}{\textbf{IP}} = Incorrect Patch, \textcolor{red}{\textbf{RS}} = Report Spam.}
\label{tab:comparison}
\renewcommand{\arraystretch}{0.75} 
\setlength{\tabcolsep}{2.0pt}
\begin{tabular}{@{} l c p{2.0cm} p{3.0cm} p{2.2cm} p{2.0cm} @{}}
\toprule
\textbf{Reference} & \textbf{Cat.} & \textbf{Failure Subtype} & \textbf{Key Reasoning Failure} & \textbf{Grounding} \\
\midrule
Ullah et al.~\cite{ullah2023can} & \textcolor{blue}{H} & Sycophantic halluc. & Hallucinates vulns in safe code to satisfy prompts & None \\
Zhang et al.~\cite{zhang2024understanding} & \textcolor{blue}{H} & Token-bias confound. & Judgment shifts with surface renaming, not logic & None \\
Chakraborty et al.~\cite{chakraborty2022deep} & \textcolor{blue}{H} & Spurious feature learn. & Learns syntactic artifacts, not exec.\ semantics & Partial \\
Pearce et al.~\cite{pearce2022asleep} & \textcolor{blue}{H} & Tool-use confab. & Agents misinterpret tool output in security tasks & Tool logs \\
\midrule
Pearce et al.~\cite{pearce2023examining} & \textcolor{orange}{IP} & Symptom-only repair & Fixes surface, misses root cause, adds regressions & Compile \\
Xia et al.~\cite{xia2023keep} & \textcolor{orange}{IP} & Repair loop drift & Plausible fixes silently break arch.\ state & Tests \\
Jimenez et al.~\cite{jimenez2024swe} & \textcolor{orange}{IP} & Cross-file invariant loss & Cannot maintain invariants across files & Tests \\
Pearce et al.~\cite{pearce2022asleep} & \textcolor{orange}{IP} & Security regress.\ intro. & Generated patches embed new CWEs & Static analysis \\
\midrule
Kabir et al.~\cite{kabir2024who} & \textcolor{red}{RS} & Fluency-as-credibility & Coherent structure masks technical errors & User study \\
Stenberg~\cite{sharwood2026curl} & \textcolor{red}{RS} & Automated flooding & Plausible narratives without PoC validation & Triage logs \\
Dong et al.~\cite{11068801} & \textcolor{red}{RS} & Undisclosed-fix repackage & Silent fixes provide undetectable raw material & Commit diffs \\
\bottomrule
\end{tabular}
\end{table}

\subsection{Semantic Repackaging and Report Spam}
\label{subsec:semantic_repackaging}

The third branch emerges from deliberate use. Language models can reshape existing information into forms that appear original and authoritative. Legacy vulnerabilities are rewritten, minor issues inflated, and large volumes of reports generated with minimal grounding.

Kabir et al. (2024) showed that readers consistently prefer AI-generated explanations due to clarity and structure, even when those explanations are incorrect~\cite{kabir2024who}. Fluency becomes a proxy for credibility, and what once served as a useful heuristic in triage becomes a point of exploitation. Stenberg documented the impact on \textit{cURL}, where automated submissions flood the pipeline with well-structured but unverifiable reports lacking functional proofs of concept~\cite{sharwood2026curl}.

The silent vulnerability fixes documented by Dong et al. (2025) deepen this picture~\cite{11068801}. A substantial portion of real remediation never enters the public record. An LLM that has ingested commit diffs from open-source repositories may ``discover'' a vulnerability that was silently fixed years ago, package it as a novel finding, and submit it to a bug bounty platform. Because no CVE was assigned, there is no public contradiction. The slop is undetectable by any lookup-based validation. Autonomous agent frameworks compound this by generating reports at scale. When pointed at static analyzer output (which itself has high false-positive rates), the result is a cascade: low-confidence alerts are laundered through fluent language into reports that appear analytically rigorous despite the underlying signal being noise.

\subsection{Summary and Comparative Analysis}
\label{subsec:comparative_table}

Figure~\ref{fig:Taxonomy} and Table~\ref{tab:comparison} converge on a single conclusion. The manifestations differ, but the origin remains the same. Hallucinated vulnerabilities, incorrect patches, and semantic repackaging all arise when probabilistic generation stands in for causal verification. The ``Grounding'' column is particularly telling: not a single study provides execution-level validation of claims in LLM outputs. The strongest grounding is compile checking or static analysis, both operating far below what security reasoning demands. Until generation is anchored to verification, these categories will continue to evolve, becoming more fluent and more difficult to distinguish from genuine analysis.

%% file: tex/4_discussion.tex
\section{The Empirical Divide: Human Deductive Reasoning versus AI Probabilistic Generation}
\label{sec:reasoning_divide}

\begin{figure}
    \centering
    \includegraphics[width=1\linewidth]{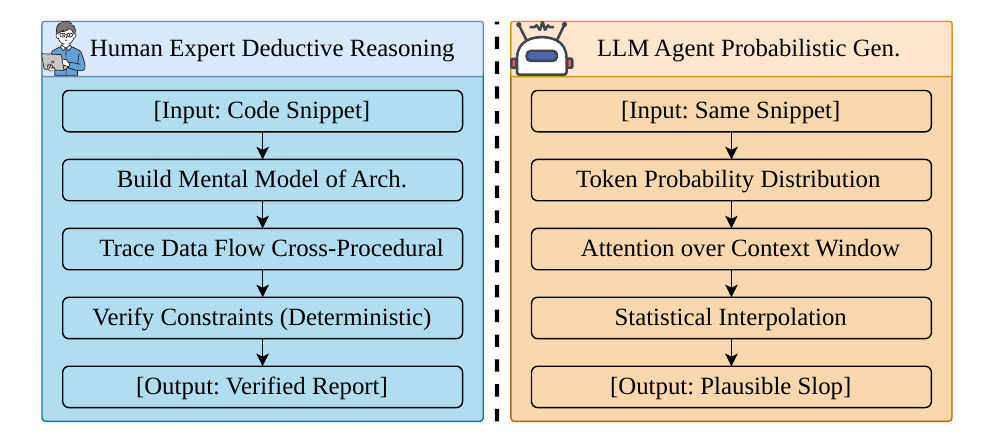}
    \caption{Epistemic gap between human analysts and AR LLMs. Dashed arrows on the LLM side indicate optional external interactions (tool use, retrieval) that partially mitigate but do not close the gap. The human side includes heuristic shortcuts alongside deductive steps.}
    \label{fig:gap}
\end{figure}

Figure~\ref{fig:gap} captures a tension at the heart of modern vulnerability analysis. Two systems receive the same input and walk in fundamentally different directions. One builds, tests, and verifies. The other predicts, connects, and completes. The gap is not a matter of performance. It is a difference in how each system understands what it means to be correct.

\subsection{Human Cognitive Models in Vulnerability Assessment}
\label{subsec:human_cognition}

On the left side of Figure~\ref{fig:gap}, the human analyst constructs a mental model of the system. This model evolves as the analyst moves through the code, tracing data across functions, resolving dependencies, and identifying constraints governing execution. Votipka et al. observed that this process is a constant negotiation between architectural understanding and instruction-level detail~\cite{votipka2020observational}. The analyst forms a hypothesis, follows it through the program, and abandons it when it conflicts with observed behavior. Reachability is not assumed. It is proven.

This mode of reasoning depends on a persistent internal state carrying forward invariants never explicitly written yet always enforced. When a potential vulnerability appears, it is treated with skepticism. The question is not whether it looks like a known pattern but whether it survives the full chain of execution without contradiction.

It would be a mistake to idealize this process. Human analysts are not uniformly deductive. Real-world triage does not operate under ideal conditions. Experienced researchers rely heavily on pattern recognition and heuristic pruning, recognizing that a code structure ``looks like'' a known vulnerability class without fully tracing every path. They operate under time pressure, incomplete information, and cognitive biases such as anchoring on the first plausible hypothesis. Votipka et al. note that analysts frequently skip verification steps when a finding ``feels right,'' a behavior that produces its own false positives. The difference is one of degree rather than kind. Human heuristics are grounded in experience with real execution failures. Model heuristics are grounded in token co-occurrence statistics with no causal relationship to program behavior. When humans err, they err within a framework that admits the possibility of error. When models err, they lack that meta-cognitive capacity.

\subsection{Architectural Constraints of AR Generation}
\label{subsec:ai_architecture}

On the right side of Figure~\ref{fig:gap}, LLMs follow a different path. They do not construct an internal execution trace. They process input as a sequence of tokens, estimating the probability of what should come next based on patterns learned during training. The notion of likelihood replaces the notion of correctness.

This distinction becomes pronounced as task complexity increases. Liu et al. (2024) showed that when context grows, attention becomes unevenly distributed, leading to a ``lost in the middle'' effect where critical information fades~\cite{liu2024lost}. For vulnerability analysis, long-range dependencies and cross-procedural relationships are not consistently preserved. The model samples fragments of context and assembles them into a coherent narrative. What appears as reasoning is often reconstruction. Variable names, API calls, and security patterns are stitched together because they co-occur in similar contexts, not because they have been verified within a single execution path. When gaps emerge, they are filled with statistically plausible continuations. A legitimate data flow can quietly transition into an invented function call, not as an explicit error but as a natural extension of the learned distribution.

Increasing model scale does not resolve this tension. It often amplifies it, allowing the system to generate more coherent and persuasive narratives while remaining anchored to the same probabilistic foundation.

\subsection{Partial Bridges: Chain-of-Thought, Tool Use, and Their Limits}
\label{subsec:partial_bridges}

The picture in Figure~\ref{fig:gap} would be incomplete without acknowledging architectural modifications designed to narrow this gap. Chain-of-thought (CoT) prompting, tool-augmented generation, and retrieval-grounded approaches each inject structure into otherwise unconstrained token prediction. The question is whether they help enough.

CoT encourages the model to externalize intermediate reasoning steps. In vulnerability analysis, this can manifest as explicit ``trace data from source to sink'' instructions. The improvement is concentrated in simple, single-function vulnerabilities where the reasoning chain is short. On cross-procedural vulnerabilities requiring three or more hops, CoT's benefit diminishes sharply. In several documented cases the model produced longer but equally incorrect reasoning chains, more words with no more verification. The model narrates its way toward an answer it has already probabilistically committed to.

Tool-using agents introduce a more promising mechanism. Frameworks such as ReAct allow models to invoke external utilities during generation. In practice, the results are mixed. The agent's ability to correctly formulate the query remains subject to the same token-level biases that produce slop in standalone generation. Pearce et al. observed that AI agents tasked with autonomous security auditing frequently invoke tools incorrectly, passing wrong arguments, querying the wrong functions, or misinterpreting structured output~\cite{pearce2022asleep}. The tool call becomes a prop in a narrative rather than a genuine constraint on reasoning.

Retrieval-augmented generation (RAG) anchors the model's output to retrieved documents. This can prevent fabrication of nonexistent CVEs. The limitation is that retrieval operates at the document level, not the execution level. A retrieved CVE description confirms that a vulnerability class exists. It does not confirm that the specific code under analysis instantiates that class. RAG reduces factual hallucination but does not address logical hallucination, the case where the model draws a valid-sounding but causally incorrect inference from accurately retrieved facts.

These modifications narrow the gap but do not close it. They add friction to the generation process, which helps, but they do not substitute for execution-grounded verification.

\subsection{Operationalizing the Gap: The Deductive Coverage Score}
\label{subsec:dcs}

To move beyond schematic arguments, the reasoning gap must be measurable. We propose the \textbf{Deductive Coverage Score (DCS)} as a proxy for the degree to which a generated vulnerability claim is grounded in explicit, verifiable evidence rather than statistical interpolation.

Given a vulnerability claim $C$, a domain expert (or deterministic verification tool) decomposes $C$ into $k$ atomic constraints $\{c_1, c_2, \ldots, c_k\}$ that must each hold for the claim to be valid. For a typical memory-corruption claim, these might include: $c_1$ (tainted source reachable from a public entry point), $c_2$ (data flows to a dangerous sink without effective sanitization), $c_3$ (sink operation triggerable under realistic preconditions), $c_4$ (target buffer size insufficient for expected input). The DCS is:

\begin{equation}
\text{DCS}(C) = \frac{1}{k} \sum_{i=1}^{k} \mathbf{1}[\text{claim } c_i \text{ is explicitly grounded in the output}]
\end{equation}

``Explicitly grounded'' means the output contains a concrete evidence anchor for $c_i$: a specific code reference, execution trace, concrete input specification, or formal proof fragment. Vague appeals to ``the data flows through several functions'' do not count.

%% file: tex/5_discussion2.tex
\section{Mitigating AI Slop: Content Provenance and Verification Strategies}
\label{sec:mitigation_strategies}

Before surveying mitigation approaches, it is necessary to clarify what we are trying to mitigate. The core failure of AI slop is not that it was produced by a machine. A human analyst working from incomplete information under time pressure can produce an equally hollow report. An LLM-assisted analysis that has passed formal verification is useful regardless of its origin. The variable that matters is not provenance but correctness: whether the artifact carries verifiable substance that survives independent validation.

Much of the initial mitigation effort targeted provenance rather than correctness. This is understandable as a first response, but as we show, it encounters fundamental limits in security domains. We organize the landscape into three phases: historically attempted but misaligned approaches, ecosystem-level attestation, and the verification-first paradigm.

\subsection{The Misalignment of Passive Statistical Detection}
\label{subsec:passive_detection}

The initial wave of AI mitigation relied on statistical analysis of generated text to identify machine provenance. DetectGPT by Mitchell et al. (2023) identifies machine-generated content by evaluating probability curvature of a text sequence~\cite{mitchell2023detectgpt}. The premise is that LLMs generate text in negative curvature regions of the log-probability function, producing sequences with lower perplexity and burstiness than human writing. Supervised classifiers fine-tuned on architectures such as RoBERTa were deployed to detect semantic anomalies.

These approaches degrade significantly when applied to vulnerability reports, and the failure mode reveals why provenance is the wrong target. Security artifacts are inherently formulaic with low lexical diversity. Human-written bug bounty reports rely on rigid templates, structured stack traces, and standardized formatting that naturally produce the low perplexity scores statistical detectors flag as artificial. The detector distinguishes fluent-from-template from fluent-from-model, a distinction with no bearing on whether the underlying claim is true.

Sadasivan et al. (2025) formalized this fragility, proving that as generative model output distributions approximate human writing distributions, the ROC AUC for any statistical detector converges to random chance~\cite{sadasivan2023can}. Applying these detectors to triage platforms produces unacceptably high false-positive rates, penalizing legitimate researchers while remaining vulnerable to adversarially engineered burstiness. Passive detection asks ``was this written by a human?'' when it should ask ``is this claim correct?''

\subsection{Cryptographic Watermarking and Ecosystem Attestation}
\label{subsec:cryptographic_attestation}

Kirchenbauer et al. (2023) proposed soft watermarking that alters vocabulary distribution during token selection, dividing the token space into permissible and restricted lists to verify provenance~\cite{kirchenbauer2023watermark}. In code generation and security patching, the vocabulary space is extremely constrained. Forcing a model to select a watermarked token in a deterministic logic chain frequently produces an incorrect API call, an invalid variable name, or a broken payload. Watermarking does not just fail to detect slop. It can actively produce slop by corrupting the logic it is supposed to authenticate.

Recognizing this, the community is pivoting toward ecosystem-level attestation. Inspired by supply chain frameworks such as \textit{in-toto}~\cite{torres2019toto}, platforms are moving toward proving human provenance rather than machine provenance. Requiring researchers to cryptographically sign proofs-of-concept and execution traces using authenticated public key infrastructure shifts the verification burden. Attestation is a pragmatic infrastructure defense, but it remains a provenance mechanism. A cryptographically signed report can still describe a phantom exploit path. The signature proves who submitted it, not whether it is true.

\begin{figure}
    \centering
    \includegraphics[width=1\linewidth]{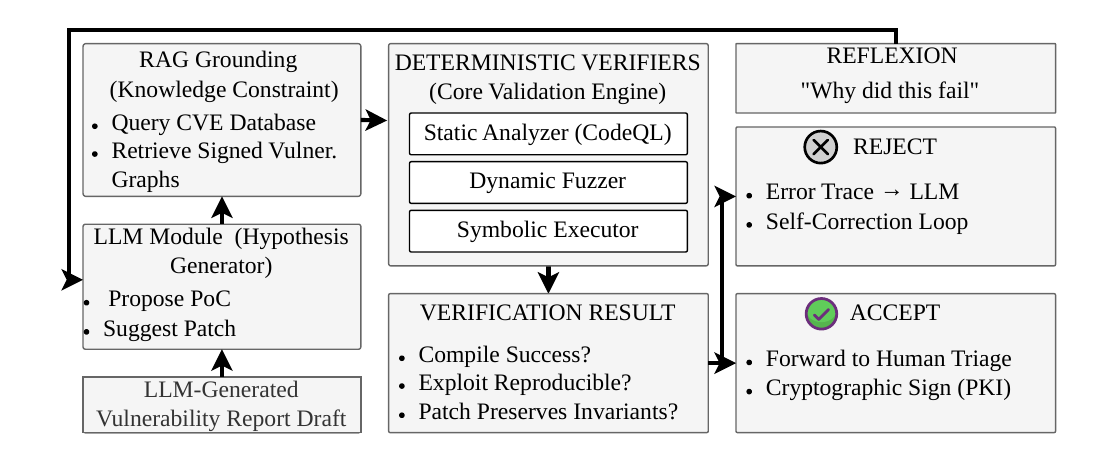}
    \caption{Neuro-symbolic verification architecture for trustworthy triage. Each component maps to prior systems and documented limits discussed in Section~\ref{subsec:active_verification}.}
    \label{fig:placeholder}
\end{figure}

\subsection{Active Verification and Neuro-Symbolic Architectures}
\label{subsec:active_verification}
Given the limitations of passive detection and watermarking, the most resilient strategy shifts the target from provenance to correctness. The architecture in Figure~\ref{fig:placeholder} embodies a different premise: the language model generates hypotheses, and a pipeline subjects each hypothesis to deterministic verification before it reaches a human analyst. Below we map each component to concrete prior systems and their documented limits on security inputs.

\textbf{Retrieval-Augmented Grounding.} The model anchors claims in externally verified sources such as CVE databases~\cite{lewis2020retrieval}. RAG operates at the document level rather than the execution level. A retrieved CVE entry confirms a vulnerability class exists but does not confirm that the specific code under analysis instantiates it. Lewis et al. note that retrieval noise can degrade generation quality below the parameter-only baseline~\cite{lewis2020retrieval}. RAG reduces factual hallucination but does not address logical hallucination.

\textbf{Static Analysis Verification (CodeQL).} Generated claims are handed to an engine operating on the actual AST and control-flow graph. CodeQL is widely deployed and integrated into GitHub's infrastructure~\cite{avgustinov2016ql}. Its coverage on security-critical bug classes is uneven. Memory-safety vulnerabilities involving complex heap layouts require custom queries, and its path-sensitive analysis does not model heap state by default. Template-heavy C++ and indirect calls through vtables produce false negatives that undermine verification~\cite{10.1145/3533767.3534380}. CodeQL can confirm certain claims but cannot refute complex ones with reliable completeness.

\textbf{Dynamic Verification (Fuzzing).} For claims involving exploitability, the generated proof-of-concept is compiled and executed under a coverage-guided fuzzer such as AFL or libFuzzer~\cite{manes2019art}. Coverage-guided fuzzers struggle with magic-byte comparisons, deep conditional branches, and path constraints requiring specific heap layouts~\cite{li2017steelix}. Furthermore, the verification quality depends entirely on the LLM-generated harness. As documented in Section~\ref{subsec:incorrect_patches}, LLM-generated code frequently contains errors that prevent valid harnesses from compiling or exercising the intended path~\cite{zhang2024how}.

\textbf{Symbolic Execution Verification.} A symbolic executor such as KLEE~\cite{klee} or angr~\cite{ager} explores the path space to determine whether claimed constraints are satisfiable. This approach provides strong theoretical guarantees but faces path explosion as feasible paths grow exponentially with branch count. Constraint solvers time out on non-linear arithmetic, string operations, and floating-point comparisons. These limitations restrict practical coverage on the complex programs where slop is most likely to appear.

\textbf{Iterative Self-Correction (Reflexion).} When a claim fails verification, the error trace is fed back to the model. Shinn et al. (2025) introduced Reflexion as a framework for verbal reinforcement learning through iterative external feedback~\cite{shinn2024reflexion}. Effectiveness depends heavily on the granularity of this feedback. There is no theoretical guarantee of convergence. Huang et al. provide evidence that without access to external ground truth, models struggle to correct their own reasoning even when explicitly presented with their errors~\cite{huang2023large}. In security contexts, where error traces from fuzzers or symbolic executors can be voluminous and partially irrelevant, this limitation poses a tangible risk of reinforcing rather than resolving invalid generation.

\textbf{Synthesis.} Each component has blind spots. The pipeline raises the floor by requiring a hallucinated claim to survive at least two independent verification stages before reaching a human. Most slop fails at the first hurdle. The architecture shifts the burden from the human analyst to the system and ensures human judgment is exercised only on claims that have survived deterministic tests.

%% file: tex/6_future.tex
\section{Open Challenges and Future Directions}
\label{sec:challenges}

Active verification architectures provide a foundational defense, but the rapid evolution of foundation models continuously shifts the threat landscape. We outline three trajectories, with concrete specifications for the evaluation instruments the field currently lacks.

\begin{table}[!htbp]
\tiny
\centering
\caption{Existing benchmarks vs.\ proposed evaluation instruments.}
\label{tab:benchmarks}
\begin{tabularx}{\textwidth}{@{} l X c X @{}}
\toprule
\textbf{Benchmark} & \textbf{Focus} & \textbf{Security} & \textbf{Gap for AI Slop} \\
\midrule
\textbf{HumanEval / MBPP} & Standalone code synthesis & None & Binary pass/fail; no security context \\
\textbf{SWE-bench}~\cite{jimenez2024swe} & Repo-level patching & Partial & Evaluates regression, ignores deceptive fluency \\
\textbf{CyberSecEval}~\cite{bhatt2023cyberseceval} & Compliance \& insecure coding & High & Tests jailbreaks, not benign-code hallucination \\
\midrule
\textbf{CVE-Bench} & Phantom exploit detection & High & Isolates deceptive generation with execution-validated ground truth \\
\textbf{Slop-Score} & Fluency vs.\ substance & High & Continuous metric penalizing the fluency-substance gap \\
\bottomrule
\end{tabularx}
\end{table}

\subsection{Standardizing Benchmarks for Deceptive Generation}
\label{subsec:future_benchmarks}

Existing benchmarks reward models for appearing correct rather than being verifiably right. HumanEval and MBPP reduce evaluation to binary functional outcomes. SWE-bench considers repository context but overlooks whether the reasoning behind a patch is sound~\cite{jimenez2024swe}. CyberSecEval asks whether a model will comply with malicious intent but not what happens when the intent is benign and the model invents a vulnerability anyway~\cite{bhatt2023cyberseceval}. We specify two instruments to address this gap.

\subsubsection{CVE-Bench}
CVE-Bench tests whether triage systems can separate grounded vulnerability reports from fluent outputs containing phantom exploit paths. Positive samples consist of verified recent CVEs paired with reproducible proofs-of-concept. Negative samples are LLM-generated reports on the same codebases, filtered to retain only those citing nonexistent CVEs or fabricating paths while scoring high on human fluency ratings.

The benchmark covers three tasks: binary classification (F1 on phantoms), ungrounded claim localization, and triage ranking (nDCG). To prevent the evaluation from collapsing into simple retrieval, the design uses temporal holdouts, a no-identifier variant forcing technical-content judgment, and mechanically paraphrased positives. Expected baselines span raw LLMs, chain-of-thought prompting, retrieval-augmented generation, and the full neuro-symbolic pipeline in Section~\ref{subsec:active_verification}.

\subsubsection{Slop-Score}
Slop-Score provides a continuous metric quantifying the gap between linguistic polish and verifiable substance in a security artifact $O$:
\begin{equation}
S(O) = \frac{F(O)}{\alpha \cdot D(O) + \beta \cdot C(O) + \epsilon}
\end{equation}
 $F(O)$ is normalized fluency, computed via reference-model perplexity against a calibration corpus of human reports. $D(O)$ is evidence density, the fraction of technical claims in $O$ backed by explicit anchors such as code references, line numbers, or execution traces. $C(O)$ is the constraint satisfaction ratio, the fraction of required exploit-path constraints for the claimed CWE class that are explicitly verified in the output. $\alpha$ and $\beta$ are tunable hyperparameters prioritizing constraint verification over mere evidence citation, and $\epsilon$ prevents division by zero. Higher $S$ indicates fluent but hollow output. The metric is decomposable: a high score directs attention to whether the failure is driven by missing evidence, unverified constraints, or unusually polished language. Validation requires computing the score on a held-out set of expert-written versus LLM-generated phantom reports, measuring discrimination via AUC-ROC, and ablating each component to confirm its contribution.

\subsection{Defending Against the Adversarial Weaponization of Slop}
\label{subsec:adversarial_slop}

Any filtering mechanism invites adversaries to study its boundaries. Open-weight models lower the barrier to fine-tuning specialized generators designed to evade detection. Zou et al. (2023) demonstrated how fragile model alignment is under targeted pressure, where constructed inputs force specific harmful outputs~\cite{zou2023universal}. In vulnerability triage, this fragility takes a new form. Adversarial slop can be engineered to exploit assumptions of verification pipelines, trigger parsing edge cases, or consume disproportionate computational resources.

A particularly insidious variant exploits the component-level gaps documented in Section~\ref{subsec:active_verification}. An adversary aware that CodeQL has poor heap-coverage, that symbolic execution times out on complex loops, and that fuzzers struggle with magic-byte comparisons can craft a phantom report designed to pass through these blind spots. The report would describe a use-after-free with plausible but unverifiable heap assumptions, knowing no single component can conclusively refute it. Defending against this requires pipelines that anticipate manipulation. Triage architectures must account for computational cost, identifying inputs designed to exhaust resources and deprioritizing them before the verification budget is spent.

\subsection{Verifiable Reasoning and Human-Centric Triage}
\label{subsec:human_centric}

Current generative models lack an internal sense of contradiction. When they fail, they fail with confidence, and when asked to reflect, they tend to produce explanations reinforcing the original error~\cite{huang2023large}. A hallucinated exploit path does not disappear when questioned. It evolves into a more elaborate narrative defending its own existence.

A reliable path forward places verification at the center and redefines the human role. Systems should require models to expose the structure of their reasoning. Intermediate representations, execution traces, or formal proofs of exploitability become the primary artifacts, with natural language serving only as a secondary explanation layer. The model is judged not by how convincingly it speaks but by whether its claims can be independently validated. The Deductive Coverage Score proposed in Section~\ref{subsec:dcs} provides a concrete mechanism for this interaction. A triage interface displaying the DCS breakdown shows which constraints are grounded and which are asserted without evidence, giving the analyst a direct signal about where to direct attention. Trust is no longer inferred from fluency but earned through verifiability.

%% file: tex/7_conclusion.tex
\section{Conclusion}
\label{sec:conclusion}

The integration of LLMs into cybersecurity is reshaping vulnerability assessment. What depended on careful expert analysis is increasingly delegated to automated systems promising scale and efficiency. Beneath fluent reports and confident outputs lies a widening trust gap. Hallucinated vulnerabilities, incorrect patches, and recycled bug reports reflect a mismatch between how human experts reason and how models generate.

Human analysts build conclusions through constraint, verification, and deductive discipline, though they too rely on heuristics that can fail. Language models operate by extending patterns, predicting what is likely rather than what is true. Chain-of-thought prompting, tool augmentation, and retrieval grounding narrow this gap for simpler cases but do not close it for the multi-step reasoning that security demands. Statistical filtering and watermarking target provenance rather than correctness and face hard limits in logic-critical domains. Cryptographic attestation provides pragmatic infrastructure defense but cannot verify the substance of what it signs.

The path forward requires treating generation as the starting point of a verification pipeline, pairing generative models with deterministic evaluators while remaining explicit about where each evaluator's blind spots lie. This survey has moved the conversation past anecdote toward structure. The taxonomy, the measurable reasoning gap, the component-level analysis of verification architectures, and the concrete benchmark specifications all treat AI slop as a structural failure to be engineered out of the system. Trustworthy AI in cybersecurity will not emerge from better phrasing. It will come from systems that justify what they produce in terms that are checkable, reproducible, and grounded in formal reasoning.